\documentclass[traditabstract]{aa} 
\usepackage{amsmath}
\usepackage{graphicx}
\usepackage{txfonts}
\usepackage{natbib,twoopt}
\usepackage[breaklinks=true]{hyperref} 
\bibpunct{(}{)}{;}{a}{}{,} 
\makeatletter
\newcommandtwoopt{\citeads}[3][][]{\href{http://adsabs.harvard.edu/abs/#3}%
{\def\hyper@linkstart##1##2{}%
\let\hyper@linkend\@empty\citealp[#1][#2]{#3}}}
\newcommandtwoopt{\citepads}[3][][]{\href{http://adsabs.harvard.edu/abs/#3}%
{\def\hyper@linkstart##1##2{}%
\let\hyper@linkend\@empty\citep[#1][#2]{#3}}}
\newcommandtwoopt{\citetads}[3][][]{\href{http://adsabs.harvard.edu/abs/#3}%
{\def\hyper@linkstart##1##2{}%
\let\hyper@linkend\@empty\citet[#1][#2]{#3}}}
\newcommandtwoopt{\citeyearads}[3][][]%
{\href{http://adsabs.harvard.edu/abs/#3}
{\def\hyper@linkstart##1##2{}%
\let\hyper@linkend\@empty\citeyear[#1][#2]{#3}}}
\makeatother
\begin{document}

   \title{A strange star scenario for the formation of isolated millisecond pulsars}

   \titlerunning{Strange Star Scenario for Isolated MSPs}

   \authorrunning{L. Jiang et~al.}

   \author{Long Jiang\inst{1, 2, 3}, Na Wang \inst{1}, Wen-Cong Chen\inst{2, 3}, Xiang-Dong Li\inst{3}, Wei-Min Liu\inst{2}, Zhi-Fu Gao\inst{1}}

   \institute{Xinjiang Astronomical Observatory, CAS, Urumqi, Xinjiang 830011, China. \email{Wang: na.wang@xao.ac.cn; Chen: chenwc@pku.edu.cn; Li: lixd@nju.edu.cn}
         \and
              School of Physics and Electrical Information, Shangqiu Normal University, Shangqiu, Henan 476000, China
         \and   
             Key laboratory of Modern Astronomy and Astrophysics (Nanjing University), Ministry of Education, Nanjing 210046, China
             }
   \date{}

\abstract{According to the recycling model, neutron stars in low-mass X-ray binaries
were spun up to millisecond pulsars (MSPs), which indicates that all MSPs in the Galactic plane ought to be harbored in binaries.
However, about $20\%$ Galactic field MSPs are found to be solitary. To interpret this problem, we assume that the accreting neutron star in binaries may collapse and become a strange star when it reaches some critical mass limit. Mass loss and a weak kick induced by asymmetric collapse during the phase transition (PT) from neutron star to strange star can result in isolated MSPs.
In this work, we use a population-synthesis code to examine the PT model. The simulated results show that a kick velocity of $\sim60~{\rm km~s}^{-1}$ can produce $\sim6\times10^3$ isolated MSPs and birth rate of $\sim6.6\times10^{-7} {\rm ~yr}^{-1}$ in the Galaxy,
which is approximately in agreement with predictions from observations.
For the purpose of comparisons with future observation, we also give the mass distributions of radio and X-ray binary MSPs,
along with the delay time distribution.}

\keywords{stars: evolution -- stars: neutron -- pulsars: general}
\maketitle
\section{Introduction}
%
According to the widely accepted standard recycling model \citep{alp82, rad82, bha91},
millisecond pulsars (MSPs) which are characterized by short spin periods ($P_{\rm spin}\leq30{\rm ~ms}$)
and low surface magnetic fields ($B\sim10^8-10^9{\rm ~G}$) evolved from neutron star (NS) low-mass X-ray binaries (LMXBs) \citep{man04, lor08}.
Donor stars in LMXBs with initial orbital periods near or less than the so-called bifurcation period \citep{pyl89} always
lose their hydrogen envelope and evolve into low-mass He white dwarfs (WDs).
Considering the circularization due to the tidal interaction during the mass transfer, most MSPs should be located in binary systems with highly circularized orbits \citep{phi92},
except for those MSPs in dense globular clusters, which may form via some dynamical processes \citep{ver87, ver88}.
However, about $20\%$ of MSPs are isolated in the Galactic field, which is difficult to understand.
Since the predicted birth rate of Galactic MSPs is $\gtrsim3\times10^{-6}{\rm ~yr}^{-1}$ \citep{lor95,lyn98,fer07,sto07},
the birth rate of isolated MSPs should be $\gtrsim6\times10^{-7}{\rm ~yr}^{-1}$.

To solve this problem, \citet{van88} and  \citet{klu88} proposed
that the donor stars may have been ablated by the $\gamma$-ray and energetic particles
emitted by the MSPs just as happening in PSR B1957$+$20 \citep{fru88}.
It seems that this scenario is supported by the discoveries of MSP$+$planet binaries \citep{wol92}
and of PSR B1937$+$21, which was found to be orbited by an asteroid belt with total mass $\leq$0.05 earth mass \citep{sha13}.
The belt of B1937$+$21 was thought to be made up of the debris of its former companion which had been tidally disrupted.
However, other studies show that the evaporation timescale may be too long \citep{che13}
unless a very high evaporation efficiency ($\sim0.1$) is adopted \citep{jia16}.

Decades ago, the concept of the strange star (SS) was proposed \citep{ito70, bod71, far84, wit84, alc86, hae86}.
It was suggested that some of the pulsars may be SSs rather than NSs since the strange quark matter may be the most stable state of matter. Some researchers argued that NSs and SSs may coexist in nature and
when the central density of an NS rises above the critical density for quark deconfinement,
NS-SS phase transition (PT) may occur \citep{bom00, ber03, bom04, bom08, bom16, bha17}.
\citet{oli87} and \citet{hor88} suggest that the process of PT is gradual, lasting around $10^8{\rm ~yr}$.
However, \citet{che96} and \citet{ouy02} argue that the process might take place in a detonation mode and
the released energy is compatible with a core collapse supernova (CCSN).

In the present paper, based upon the idea that accreting NSs in LMXBs may reach some critical mass limit and transit to SSs promptly, we propose that the NS-SS PT with a kick velocity may account for the formation of isolated MSPs.
The details of the scenario and the population-synthesis code are described in sections 2 and 3, respectively.
Simulated results are given in section 4, while some discussions about this scenario are presented in section 5.
Finally, we make a brief summary in section 6.

\section{Strange Star Scenario}\label{sec:2}
At present, the detailed processes for the dissolution of baryons into their quark constituents are not well understood.
Based on the standard equation of state (EoS) of neutron-rich matter,
\citet{sta06} proposed that the critical density for quark deconfinement is $\rho_{\rm c}\sim5\rho_0$,
where $\rho_0\sim2.7\times10^{14}{\rm ~g~cm}^{-3}$ is the nuclear saturation density.
If the NS is fast-spinning, the central density might be centrifugally diluted.
The maximum mass of NS with an angle velocity $\Omega=2\pi /P$ can be expressed as \citep{har70,bay71}:
\begin{equation}
M_{\rm c}(\Omega)=M_{\rm c}(0)+\delta M(\Omega /\Omega_{\rm max})^2,
\end{equation}
where $\Omega_{\rm max}$ is the maximum angular velocity (in this work, we take $\Omega_{\rm max}=2000\pi {\rm ~rad~s^{-1}}$, e. g., the minimum spin period $P_{\rm min}=1{\rm ~ms}$ ); $M_{\rm c}(0)$ is the maximum mass of the non-rotating NS,
$\delta M$ represents the difference between $M_{\rm c}(0)$ and $M_{\rm c}$.
\citet{las96} showed that rigid rotation can increase the maximum mass of NS by a fraction of $20\%$,
while \citet{mor04} found that the fraction for a differentially rotating NS is $\leq50\%$.
\citet{hae07} predicted that the maximum baryon mass of differentially rotating NSs is $\geq50\%$ higher than that of non-rotating NS. In our simulation, we take $M_{\rm c}(0)=1.8{\rm ~M}_\odot$ according to \citet{akm98}
and $\delta M = 0.4{\rm ~M}_\odot$ , similarly to  \citet{las96}. If the mass of the NS exceeds the maximum mass, for example, $M_{\rm NS}\geq M_{\rm c}(\Omega)$, PT is assumed to take place.

In the recycling stage, the NS would accrete the material from the donor star. We adopt a description for the spin period evolution of the NS given by \citet{che00}:
\begin{equation}
\begin{split}
P={\rm max}[1.1(\frac{M-M_{\rm NS, i}}{M_{\odot}})^{-1}R_6^{-5/14}I_{45}(\frac{M}{M_{\odot}})^{-1/2},\\
1.1(\frac{M}{M_{\odot}})^{-1/2}R_6^{17/14}]~~{\rm ms},
\end{split}
\end{equation}
where $R_6$ is the radius of the NS in units of $10^6{\rm ~cm}$, and $I_{45}$ is the moment of inertia of the NS in units of $10^{45}{\rm~g}{\rm ~cm}^{2}$ ($R_6=I_{45}=1$ in our simulation); $M$ and $M_{\rm NS, i}$ are the current and initial masses of the NS, respectively. If the spin period of the NS is less than 10 ms, a MSP is assumed to form.

Some researchers studied the difference between the gravitational mass of NS and
SS (for different EoS) with the same baryon number \citep{bom00, dra07, mar17}.
They obtained the similar results: for NS with a mass of $\sim1.5~{\rm ~M_\odot}$, $M_{\rm NS}-M_{\rm SS}\approx 0.15~{\rm ~M_\odot}$\footnote
{A low value is also possible; for example, \citet{sch02} suggested
that the difference in the gravitational mass between NS and hyperon star is $\sim0.03\rm {~M_\odot}$.}.
According to their research, the mass loss ratio $(M_{\rm NS}-M_{\rm SS})/M_{\rm NS}$ during PT is about 0.1.
Assuming that this ratio is suitable for all NS-SS PT, in this work, we take
$\Delta M=M_{\rm NS}-M_{\rm SS}=0.1{M_{\rm NS}}$.

Following the study of \citet{che96} and \citet{ouy02}, we consider that the PT in the core of NS takes place quickly, as with CCSN,
and a kick velocity $V_{\rm k}$ is imparted to the newly born SS.
The orbital parameters change during PT can be solved following \citet{hil83, dew03, sha16}.
Due to long duration of mass transfer, the binary orbit before PT is assumed to be circular.
The positional angle of $V_{\rm k}$ with respect to the pre-PT orbital plane is set to be $\phi$
and the angle between $V_{\rm k}$ and the pre-PT orbital velocity $V_{\rm 0} (=(2\pi GM_{\rm 0}/P_{\rm orb, 0})^{1/3})$ is $\theta$.
The ratio between the semi-major axes before and after PT is:
\begin{equation}
\frac{a_0}{a}=2-\frac{M_{0}}{M_{0}-\Delta{M}}(1+\nu+2\nu~{\rm cos}~\theta),
\end{equation}
where $\nu=V_{\rm k}/V_{\rm 0}$,
$M_{0}$ and $P_{\rm orb, 0}$ are the total mass and the orbital period of the binary before PT, respectively.
Due to the influence of mass loss and kick, the eccentricity after PT can be written as:
\begin{equation}
\begin{split}
1-e^2=\frac{a_{\rm 0}M_{\rm 0}}{a({M_{0}-\Delta{M})}}[1+ 2\nu~{\rm cos}~\theta \\+ \nu^2({\rm cos}^2\theta + {\rm sin}^2\theta~{\rm sin}^2\phi)].
\end{split}
\end{equation}

For specific kick velocities and angles mentioned above, the PT process can disrupt the binary system and result in the birth of isolated MSPs.

\section{Population synthesis}
To study the total number and birth rate of isolated MSPs formed via NS-SS PT process in the Galaxy and the initial parameter-space of the progenitors,
we use the rapid binary star evolution (BSE) code,
which was developed by \citet{hur00, hur02}. Our main modifications of the code are as follows.

\subsection{Modification for SS}
In the original code, the types of stars were noted with 16 integer numbers ($kw$) from 0 to 15, where 13 is for NS and 14 is for BH \citep{hur00}. The maximum mass of NS is $3.0~M_{\odot}$. If the mass of a NS exceeds this limit during the accretion process, $kw$ will become 14.
Based on the description mentioned in the previous section, we introduce a new star type for SS for when the mass of NS is large enough, that is, $M_{\rm NS}\geq M_{c}(\Omega)$,
and we change the integer type number $kw$ from 13 (for NS) to 99 (for SS).

Considering the mass accretion of SS in binaries, there is a maximum mass of SS, $M_{\rm SS, MAX}$, for an object with mass beyond which it is assumed to collapse to a black hole, and $kw$ will change to 14.
\cite{gle00} studied the MIT bag model with different bag parameter $B$ and obtained a maximum mass of $\sim2.2{\rm ~M}_\odot$ and $\sim1.7{\rm ~M}_\odot$ for $B^{1/4}=140 {\rm ~MeV}$ and $B^{1/4}=160 {\rm ~MeV}$, respectively.
\cite{gan13} explored the density dependent quark mass model which was developed by \cite{dey98} and obtained an upper limit of $\sim2.0{\rm ~M}_\odot$. However, \cite{lai09} propose that a massive quark star of $\sim5{\rm ~M}_\odot$ is also stable for the EoS of Lennard-Jones quark matter. The maximum mass of SS strongly depends on EoS, hence, it is still a controversial topic. Furthermore, the spin evolution also influences the maximum mass of SS like NS.
In this work, we take a fixed value of $M_{\rm SS,MAX}=2.5~M_{\odot}$ in the standard model. To study its influence, a larger value $M_{\rm SS,MAX}=3.0~M_{\odot}$ is also used \citep{zhu13}.

\subsection{Initial input parameters}
In the simulation, the initial parameters are set following \citet{liu07} and \citet{che11}.
We assume that all stars are born in binary systems with a solar metallicity ($Z=0.02$) and circular orbits ($e=0$).
In the Galaxy, one binary with primary mass $M_1\geq0.8{\rm ~M}_\odot$ is thought to be born per year, thus,
a constant star formation rate $S=7.6085{\rm ~yr}^{-1}$ is adopted \citep{hur02}.
Using the initial mass function $f(M_1)$ given by \citet{kro93}, the mass distribution of the primary is set to $\Phi({\rm ln}M_1)=M_1f(M_1)$.
The mass distribution of the secondary is $\Phi({\rm ln}M_2)=M_2/M_1=q$, where $q$ is
the mass ratio, corresponding to a uniform distribution from 0 to 1.
The binary separation $a$ is assumed to follow a uniform distribution of ${\rm ln}~a$, that is, $\Phi({\rm ln}a)=k$ where $k=0.12328$ , following \citet{hur02}.
The input parameter space for $M_1$, $M_2$, and $a$ are set to be $0.8-80{\rm~M}_\odot$, $0.1-80{\rm~M}_\odot$,  $3-10000{\rm~R}_\odot$, respectively.
Setting ${\rm n}_X(=200)$ grid points in logarithmic space, we get
\begin{equation}
\delta {\rm ln}X=\frac{{\rm ln}X_{\rm max}-{\rm ln}X_{\rm min}}{{\rm n}_X-1}.
\end{equation}
where $X$ indicates $M_1$, $M_2$ and $a$.

\begin{table}{\center}
\footnotesize
\caption{\small{Input parameters of different models}}
\label{tab1}
\tabcolsep 13pt 
\begin{tabular}{c c c c c c}
\hline
\hline
\footnotesize
Model & $\lambda$ & $\alpha_{\rm CE}$  &$\sigma_{\rm PT}$ & $M_{\rm SS, MAX}$ & $f_{\rm acc}$ \\ \hline
A        & 0.5    & 3 & 60  &2.5   & 0.5   \\
B        & 0.5    & 1 & 60  &2.5   & 0.5      \\
C        & 0.5    & 3 & 20  &2.5   & 0.5      \\
D        & 0.5    & 3 & 100 &2.5   & 0.5     \\
E        & NJU\,* & 3 & 60  &2.5   & 0.5     \\
F        & NJU    & 1 & 60  &2.5   & 0.5      \\
G        & 0.5    & 3 & 60  &3.0   & 0.5      \\
I        & 0.5    & 3 & 60  &2.5   & 0.3   \\
J        & 0.5    & 3 & 60  &2.5   & 0.8   \\
\hline
\end{tabular}
\tablefoot{ Results from researchers of Nanjing university, \citet{xu10a,xu10b} and \citet{wang16}. See the text  for details. }
\end{table}

\subsection{Common envelope evolution}
As a result of ROLF, the binary probably enters a common envelope (CE) phase.
Because this process is very complicated and uncertain, we adopt an energy mechanism \citep{hur02}.
The binding energy of the envelope is:
\begin{equation}
E_{\rm bind}=\frac{GM_{\rm d}M_{\rm d,e}}{\lambda R_{\rm L}},
\end{equation}
where $M_{\rm d}$ and $M_{\rm d,e}$ are the total mass and the envelope mass of the donor star, respectively;
$R_{\rm L}$ is the Roche lobe radius, $\lambda(<1)$ is the binding energy parameter which denotes the mass distribution in the envelope \citep{web84, dek90}.
The parameter $\lambda$ for different stars had already been systematically calculated by
\citet{dew00}, \citet{pod03}, \citet{xu10a, xu10b} and \citet{wang16}
In this work, we adopt a fixed $\lambda=0.5$ following \citep{tou97}.

The efficiency parameter which describes the fraction of orbital energy transferred to
expel the envelope during the CE evolution is $\alpha_{\rm CE}=E_{\rm bind}/(E_{\rm orb,f}-E_{\rm orb, i})$ \citep{hur02},
where $E_{\rm orb,i}$ and $E_{\rm orb,f}$ are the initial and the final orbital energy of the core, respectively.
In our standard model, $\alpha_{\rm CE}=3$ is adopted following \citet{hur10},
while a lower value of $\alpha_{\rm CE}=1$ is also used.

\begin{table*}[t]
\footnotesize
\caption{\small{Predicted numbers and birth rates of radio/X-ray SS MSPs with various companion types for different models in the Galaxy.}}
\label{tab2}
\tabcolsep 13pt 
\begin{tabular*}{\textwidth}{c c c c c c c  c}
\hline
\hline
Model &{Phase of MSPs}  & MS             & Gaint                  & He MS/Gaint         & He WD              & CO WD        &  Isolated\\ \hline
A        & Radio   & 291    &347    &79      &30949      &887        &5878  \\
         &         & $1.5\times10^{-6}$ & $4.9\times10^{-6}$  &$1.8\times10^{-7}$ &$4.3\times10^{-6} $ &$1.6\times10^{-7}$  &$6.6\times10^{-7}$   \\
         & X-ray   & 3691   &1741   &$<1$    &10905      &858                &             \\
         &         & $1.8\times10^{-6}$ & $5.7\times10^{-6}$  &$1.1\times10^{-7}$ &$9.5\times10^{-7} $ &$1.0\times10^{-7}$  &   \\
B        & Radio   & 145    &46     &14      &3446       &99         &808  \\
         &         & $9.8\times10^{-7}$ & $1.2\times10^{-6}$  &$1.1\times10^{-7}$ &$6.3\times10^{-7}$   &$3.2\times10^{-8}$  &$8.1\times10^{-8}$   \\
         & X-ray   & 1588   &518    &$<1$    &2836       &267                &             \\
         &         & $1.2\times10^{-6}$ & $1.7\times10^{-6}$  &$8.3\times10^{-8}$ &$2.4\times10^{-7}$   &$2.7\times10^{-8}$  &   \\
C        & Radio   & 146       &236   &56     &35570     &592          &0            \\
         &         & $1.7\times10^{-6}$ & $5.3\times10^{-6}$  &$1.1\times10^{-7}$ &$4.9\times10^{-6} $ &$1.1\times10^{-7}$  &0   \\
         & X-ray   & 3614      &1895   &$<1$  &11416       &679           &              \\
         &         & $2.0\times10^{-6}$ & $6.4\times10^{-6}$  &$6.2\times10^{-8}$ &$9.9\times10^{-7} $ &$7.9\times10^{-8}$  &   \\
D        & Radio   & 283       &241   &43        &26841     &490        &19862   \\
         &         & $1.1\times10^{-6}$ & $4.0\times10^{-6}$  &$1.0\times10^{-7}$ &$3.7\times10^{-6} $ &$9.7\times10^{-8}$  &$2.1\times10^{-6}$   \\
         &X-ray    & 3555      &1609   &$<1$        &10269       &698          &            \\
         &         & $1.6\times10^{-6}$ &$4.8\times10^{-6}$   &$7.7\times10^{-8}$ &$9.0\times10^{-7}$  &$8.1\times10^{-8} $ &      \\
E        & Radio   & 9        &29    &0        &3061      &0       &852\\
         &         &$3.5\times10^{-8}$  &$4.1\times10^{-7}$   &0 &$4.1\times10^{-7}$  &0                     &$9.8\times10^{-8}$              \\
         & X-ray   & 142      &186   &0        &942           &0                      &              \\
         &         &$1.1\times10^{-7}$  &$5.1\times10^{-7}$   &0 &$1.0\times10^{-7}$  &0                     &              \\
F        & Radio   & 16       &132   &$<1$     &12124         &$<1$                   &2573\\
         &         & $1.3\times10^{-7}$ &$1.6\times10^{-6}$   &$2.9\times10^{-9}$     &$1.3\times10^{-6}$  &$2.3\times10^{-9}$  &$2.8\times10^{-7}$              \\
         & X-ray   & 116      &534   &$<1$     &631           &25                     &                 \\
         &         & $1.2\times10^{-7}$    &$1.7\times10^{-6}$&$2.9\times10^{-9}$     &$5.8\times10^{-8}$    &$2.3\times10^{-9}$       & \\
G        & Radio   &278    &629    &70      &36287      &767        &6702  \\
         &         & $1.5\times10^{-6}$ & $5.3\times10^{-6}$  &$1.7\times10^{-7}$ &$4.7\times10^{-6} $ &$1.2\times10^{-7}$  &$7.5\times10^{-7}$   \\
         & X-ray   & 3641  &1719   &$<1$    &9902      &703               &             \\
         &         & $1.7\times10^{-6}$ & $5.6\times10^{-6}$  &$1.3\times10^{-7}$ &$8.6\times10^{-7} $ &$9.2\times10^{-8}$  &   \\
I        & Radio   &68       & 24       & 40  &    2131     &   418 & 693\\
         &         &$5.2\times10^{-7}$  & $7.8\times10^{-7}$  &$9.6\times10^{-8}$   &$6.2\times10^{-7}$   &$9.1\times10^{-8}$  &$7.4\times10^{-8}$ \\
         & X-ray   &812      &154       &$<1$       & 4850    &    594 &    \\
         &         &$4.2\times10^{-7}$  & $9.8\times10^{-7}$  &$8.2\times10^{-8}$   &$4.2\times10^{-7}$   &$7.1\times10^{-8}$  & \\
J        & Radio   &3548     &   817     &   28      & 21009   &     68  &26308\\
         &         &$8.0\times10^{-6}$  & $7.7\times10^{-6}$  &$6.2\times10^{-8}$   &$4.7\times10^{-6}$   &$3.5\times10^{-8}$  &$2.4\times10^{-6}$  \\
         & X-ray   & 32837    & 8100 &  $<1$  & 23961    &    370  &  \\
         &         &$1.1\times10^{-5}$  & $1.4\times10^{-5}$  &$2.5\times10^{-8}$   &$2.1\times10^{-6}$   &$4.0\times10^{-8}$  & \\
\hline
\end{tabular*}
\end{table*}

\subsection{Kick velocity}

The kick velocity distribution during CCSN or PT, which may arise from the asymmetric collapses,
can be described by a Maxwellian distribution with one-dimensional rms $\sigma$.
\citet{hob05} made a statistical study of the proper motion of 233 pulsars.
Their study for 73 young pulsars with characteristic age less then 300 Myr indicated $\sigma_{\rm CC}=265 {\rm ~km~s}^{-1}$.
Since the one-dimensional mean speed of recycled pulsars in their study is $54(6) {\rm ~km~s}^{-1}$,
we set $\sigma_{\rm PT}=60 {\rm ~km~s}^{-1}$ in our simulation, while some lower and higher speeds are also used for comparison.

\begin{figure*}[t]
\centering
\begin{minipage}{8cm}
\includegraphics[scale=0.35, trim={120 40 25 55}]{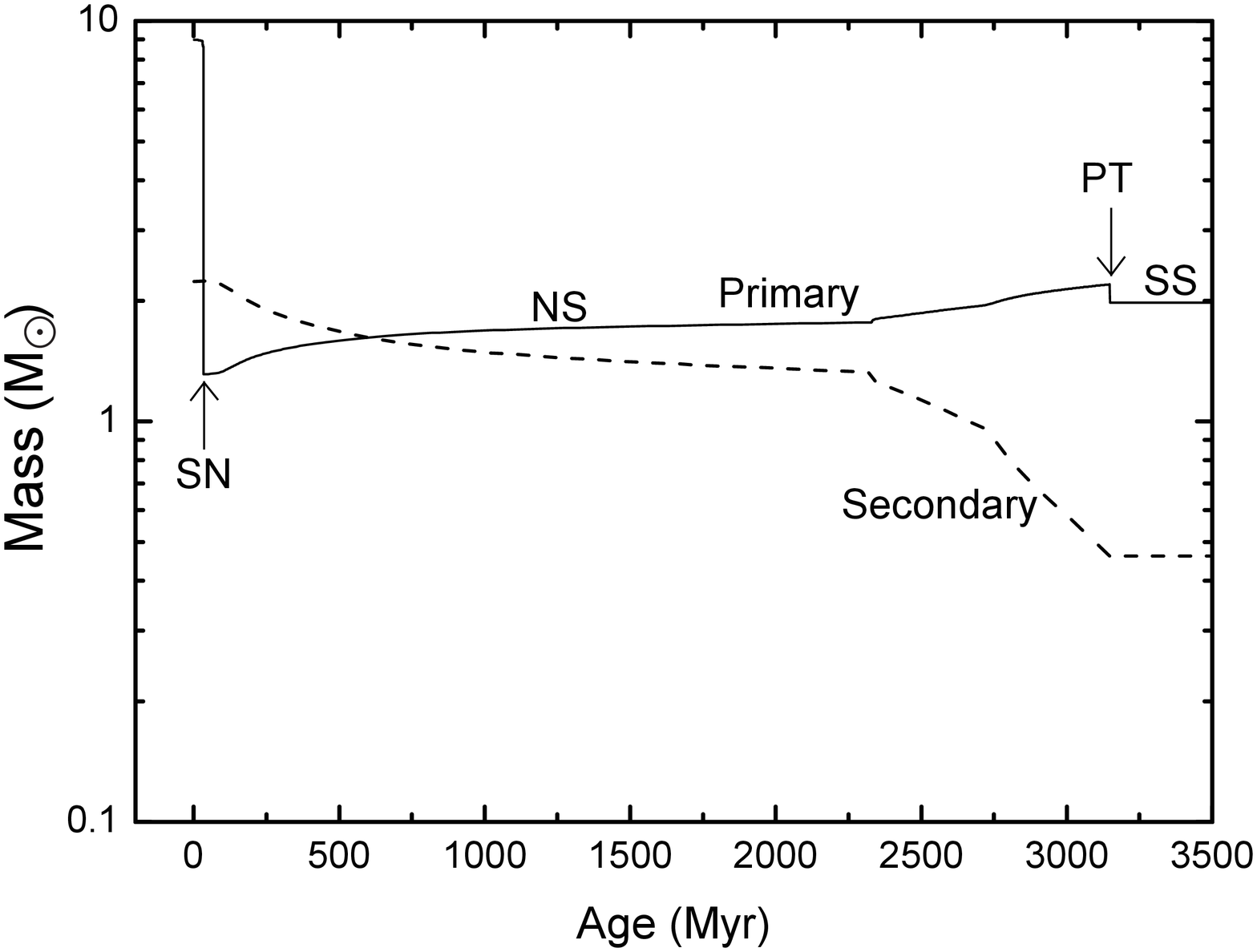}
\end{minipage}
\begin{minipage}{8cm}
\includegraphics[scale=0.35, trim={60 40 25 55}]{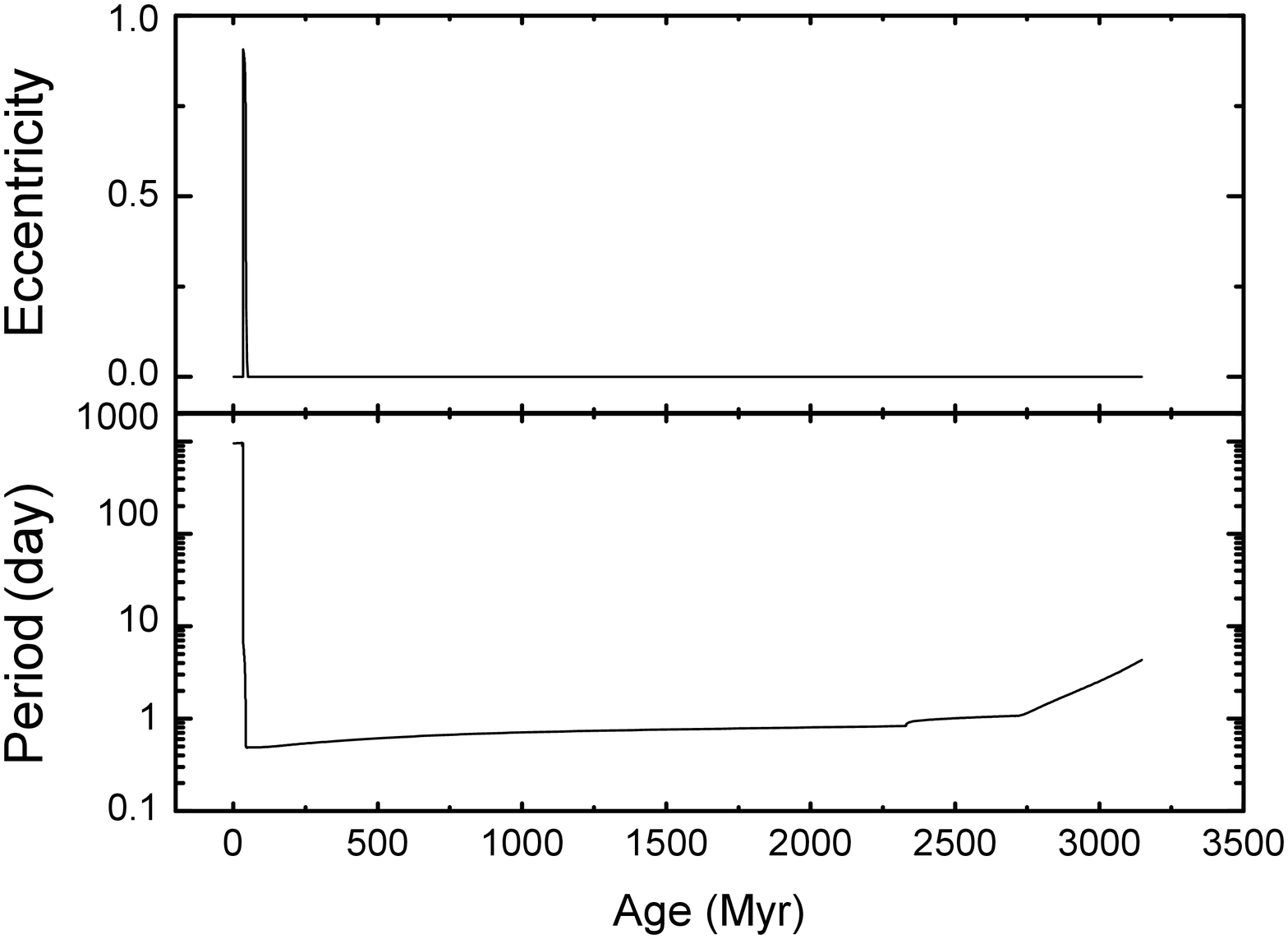}
\end{minipage}
\begin{minipage}{8cm}
\includegraphics[scale=0.35, trim={120 40 25 55}]{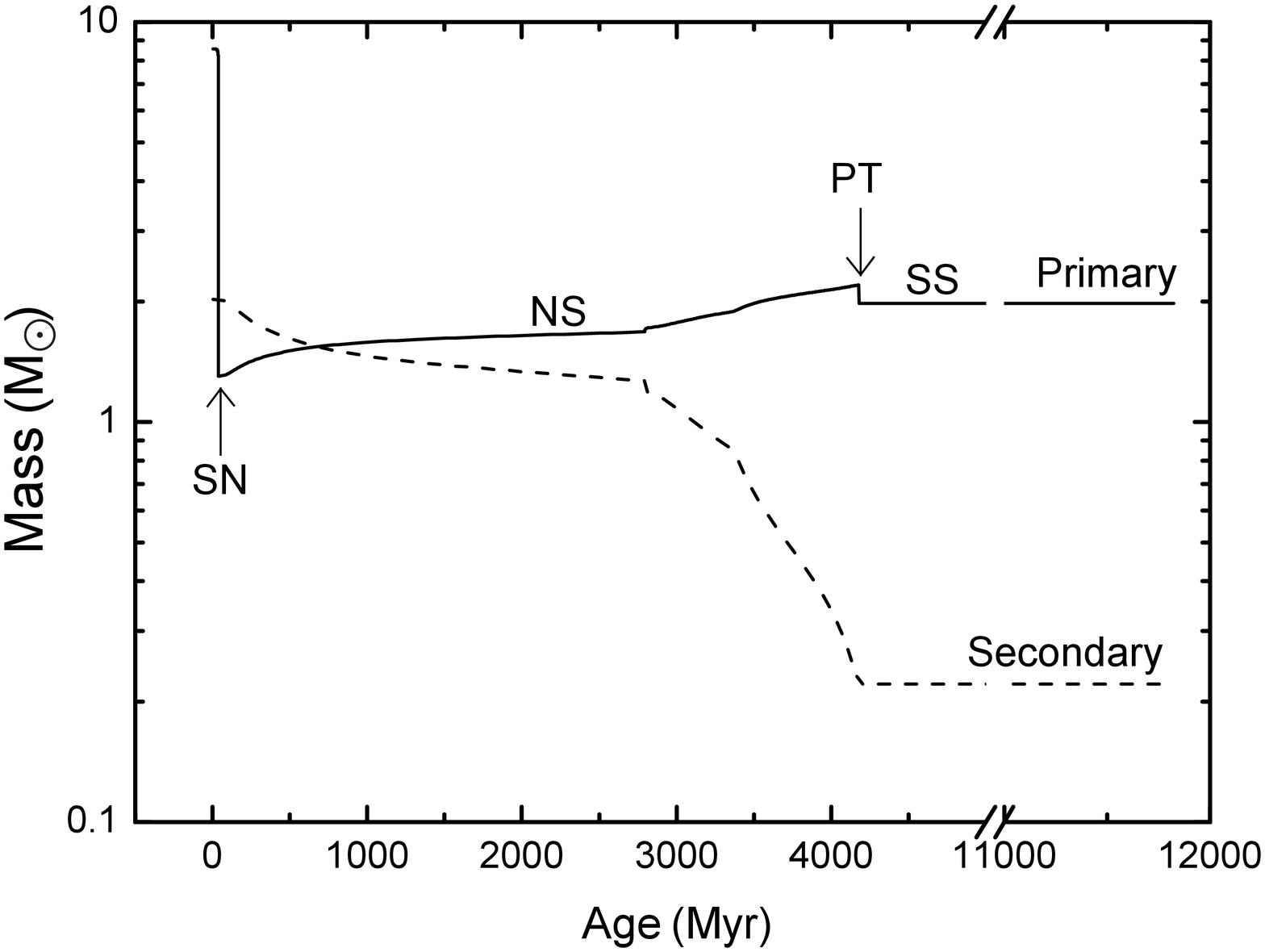}
\end{minipage}
\begin{minipage}{8cm}
\includegraphics[scale=0.35,trim={60 40 25 55}]{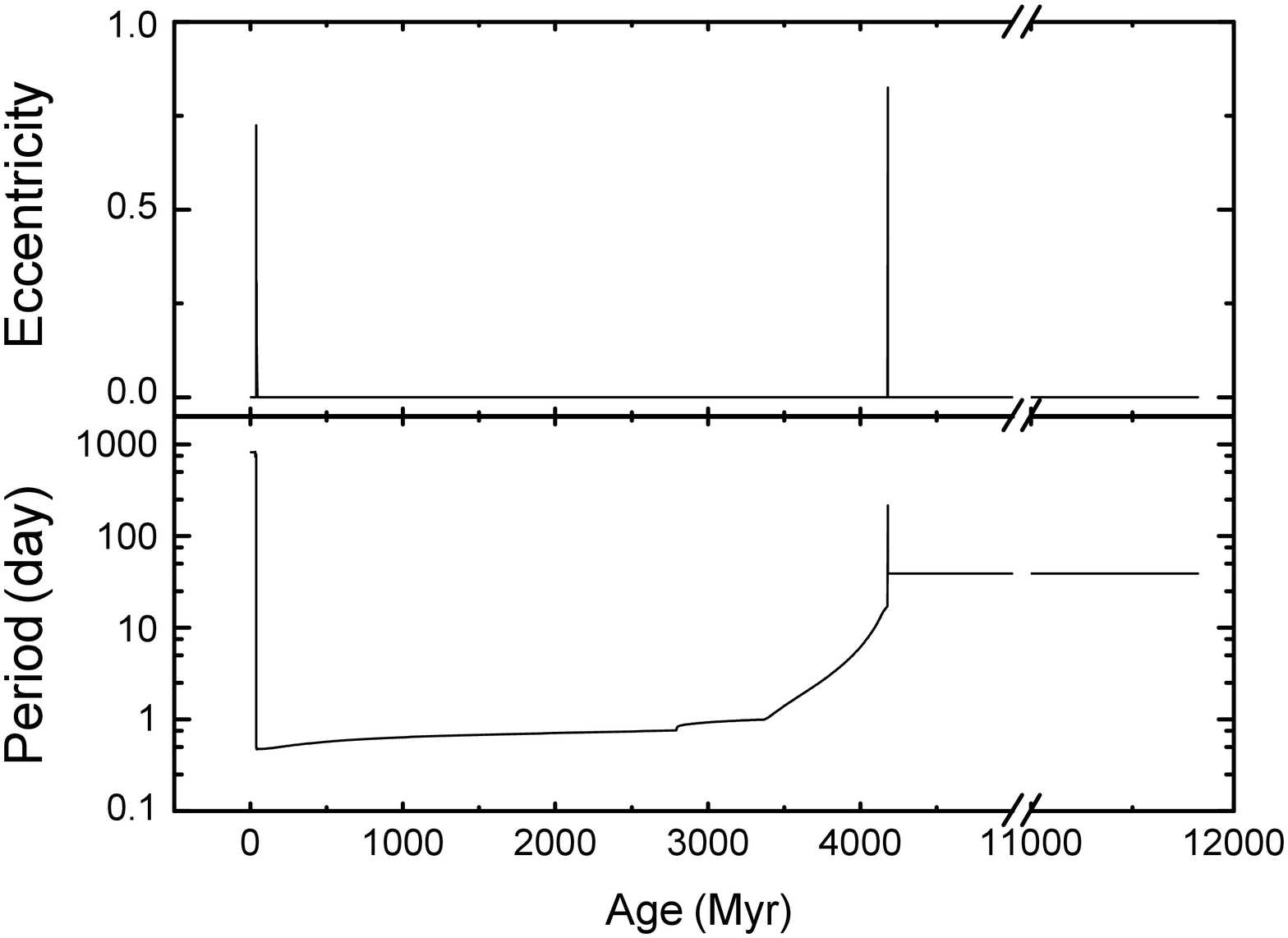}
\end{minipage}
\begin{minipage}{8cm}
\includegraphics[scale=0.35,trim={120 40 25 55}]{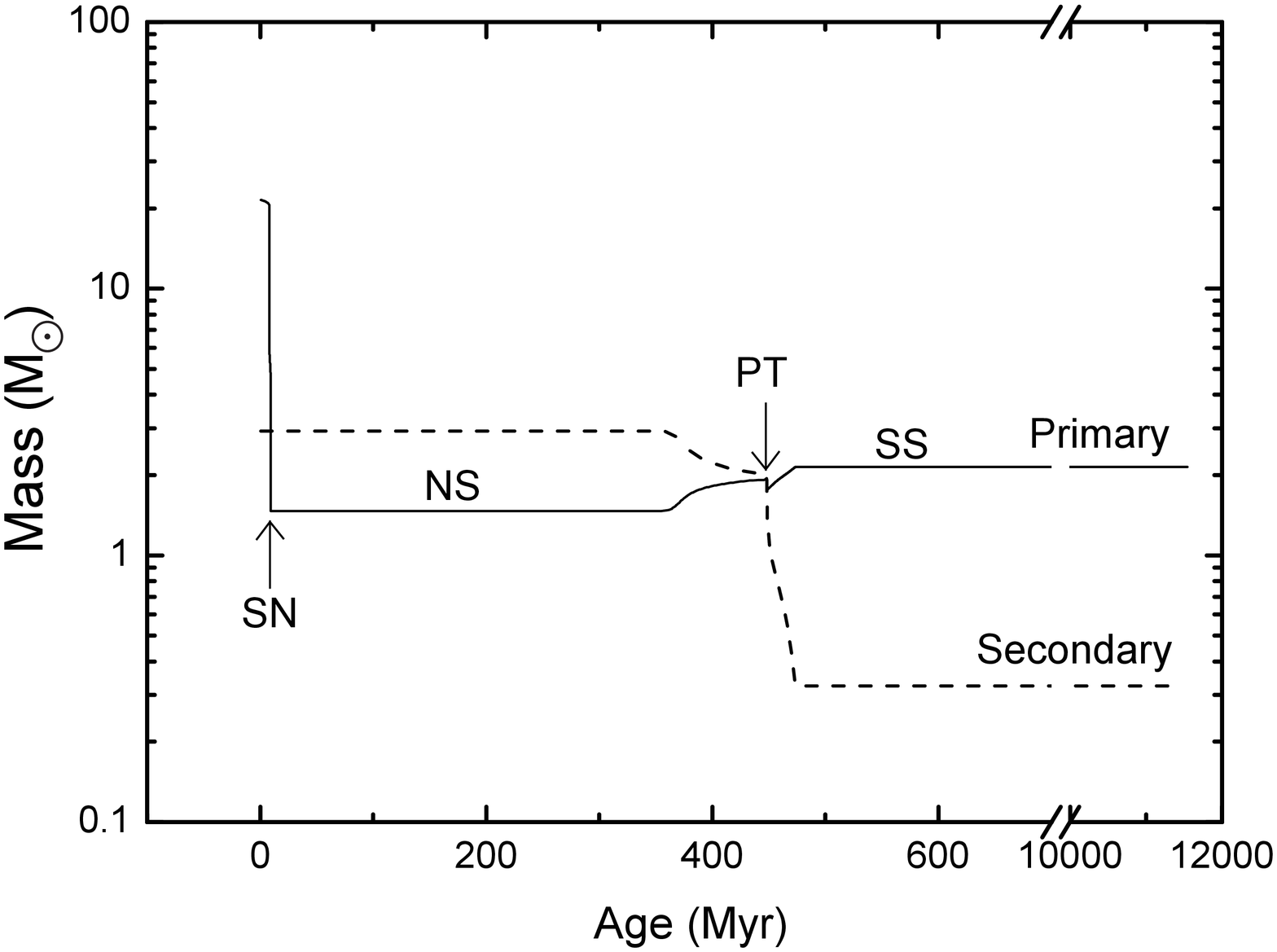}
\end{minipage}
\begin{minipage}{8cm}
\includegraphics[scale=0.35,trim={60 40 25 55}]{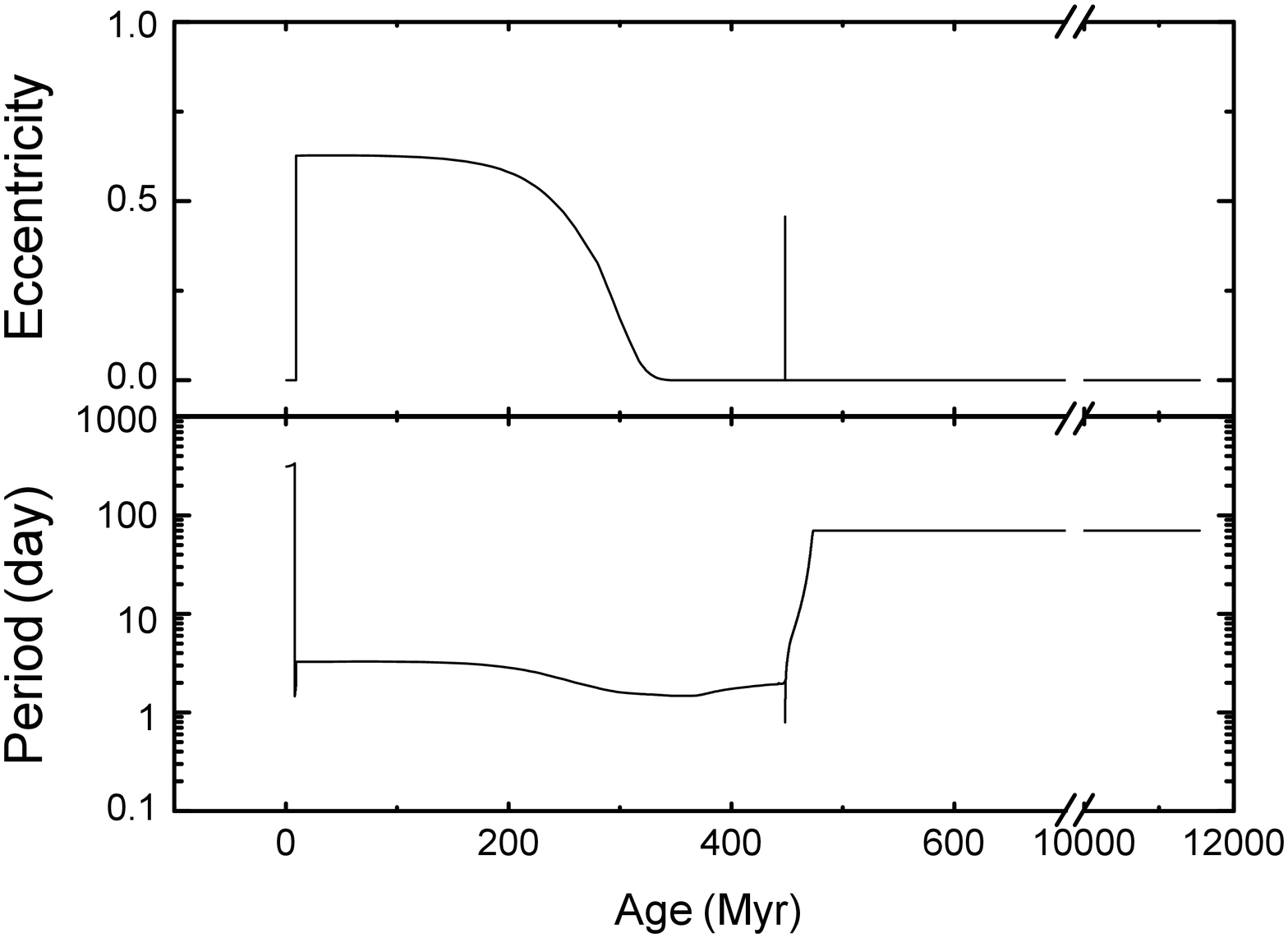}
\end{minipage}
\caption{Evolutionary traces of mass, orbital period, and eccentricity for the three cases which see the evolution into isolated MSP (top panel),
He WD$+$ SS MSP binary (middle panel), and CO WD$+$SS MSP binary (bottom panel), respectively.}
\label{fig:evolu}
\end{figure*}

\subsection{RLOF mass transfer}

After the primary evolves into an NS, the secondary will evolve to fill its Roche lobe and trigger the mass transfer. The material from the secondary is transferred to the NS at a rate of $\dot{M}_{2}$.
The accretion rate ($\dot{M}_{\rm NS}$) of the NS (also subsequent SS after PT) is thought to be limited to the so-called Eddington accretion rate $\dot{M}_{\rm Edd}$ and an accretion efficiency $f_{\rm acc}$, that is,
\begin{equation}
\dot{M}_{\rm NS}={\rm min}[\dot{M}_{\rm Edd},f_{\rm acc}|\dot{M}_2|].
\end{equation}
In our simulation, we adopt $f_{\rm acc}=0.3$, 0.8, and 0.5 \citep{pod02}. The mass loss is thought to be ejected in the vicinity of the NS in the form of isotropic winds,
carrying the specific angular momentum of the NS.

We calculate the evolution of each binary up to an age of 12 Gyr using the BSE code. During the evolution of binary systems, if a NS evolves into a SS MSP (isolated SS MSP or binary SS MSP),
it makes a contribution to the birth rate (in units of systems per year) of relevant MSPs as
\begin{equation}
\delta r=S(\Phi{\rm ln}M_1)(\Phi{\rm ln}M_2)(\Phi{\rm ln}a)\delta{\rm ln}M_1\delta{\rm ln}M_2\delta{\rm ln}a.
\end{equation}
If this kind of SS lives for a time of $\delta t$,
it makes a contribution to the number
\begin{equation}
\delta n=\delta r\delta t.
\end{equation}

Besides the assumptions mentioned above, we also consider other binary star interactions,
such as the mass transfer, accretion via stellar winds, tidal friction,
and orbital angular momentum loss via gravitational wave radiation
and magnetic braking \citep{hur02}.

\section{Simulation results}
Based on the assumptions mentioned above, we simulated the evolution of $n^{3}_{\rm X}(=8\times10^6)$
binaries.
We constructed several models with the input parameters shown in Table 1. In our standard model, Model A, $\lambda = 0.5$, $\alpha_{\rm CE} = 3$,
$\sigma_{\rm PT}=60 {\rm ~km~s}^{-1}$, $M_{\rm SS, MAX}=2.5~M_{\odot}$ and $f_{\rm acc}=0.5$.

\begin{figure*}[t]
\centering
\begin{minipage}{18cm}
\includegraphics[scale=0.7,trim={20 40 100 40}]{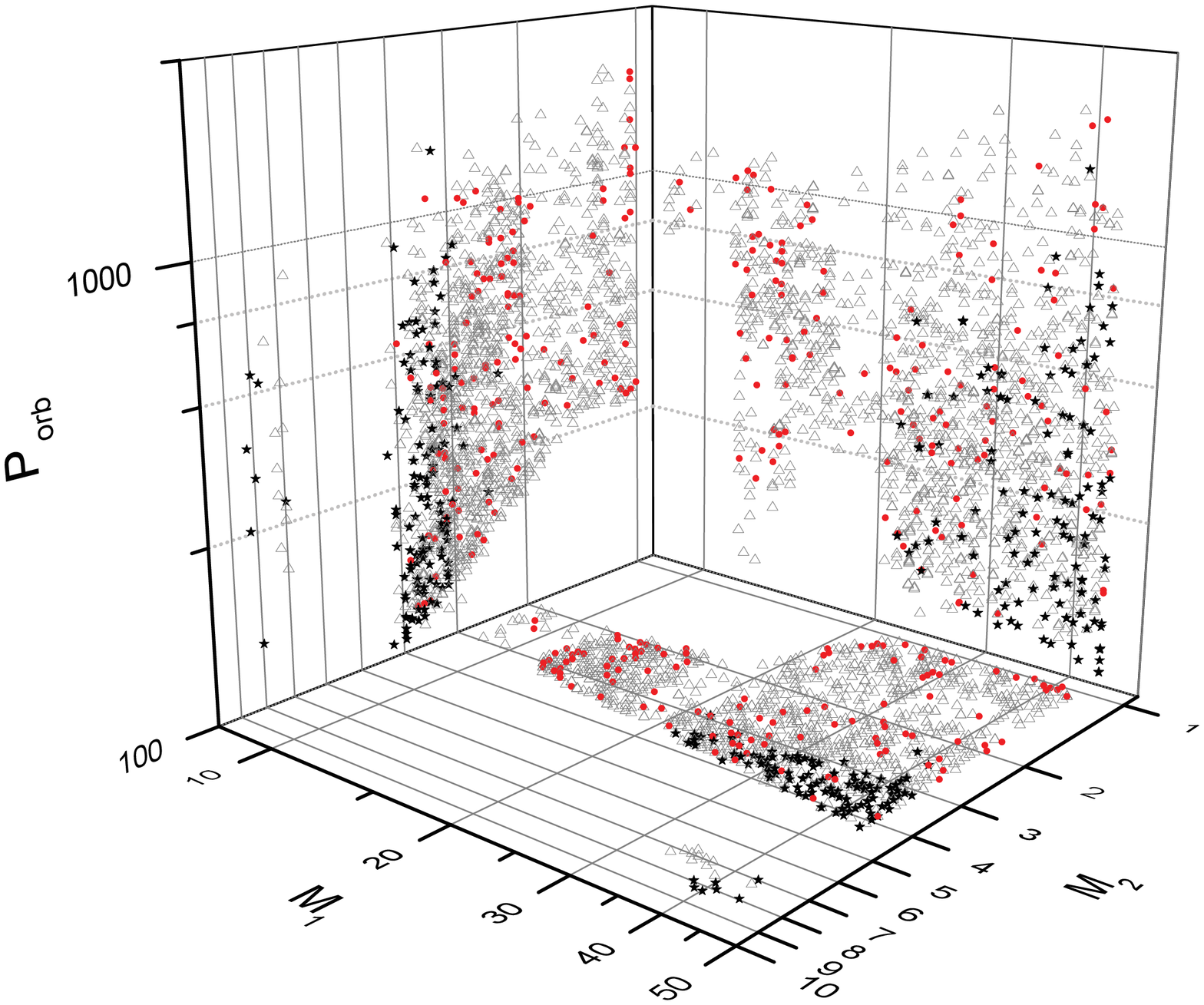}
\caption{Projections of the 3D-distribution of the initial primary masses ($M_1$),
the secondary masses ($M_2$), and orbital periods ($P_{\rm orb}$)
of primordial binary systems that would evolve into isolated MSPs (red solid circles),
He WD$+$SS binary MSPs (grey open triangles), and CO WD$+$ SS binary MSPs (black solid stars), respectively.}
\label{fig:IMMP3D}
\end{minipage}
\end{figure*}

\begin{figure}
\centering
\includegraphics[scale=0.34,trim={20 30 0 0},angle=0]{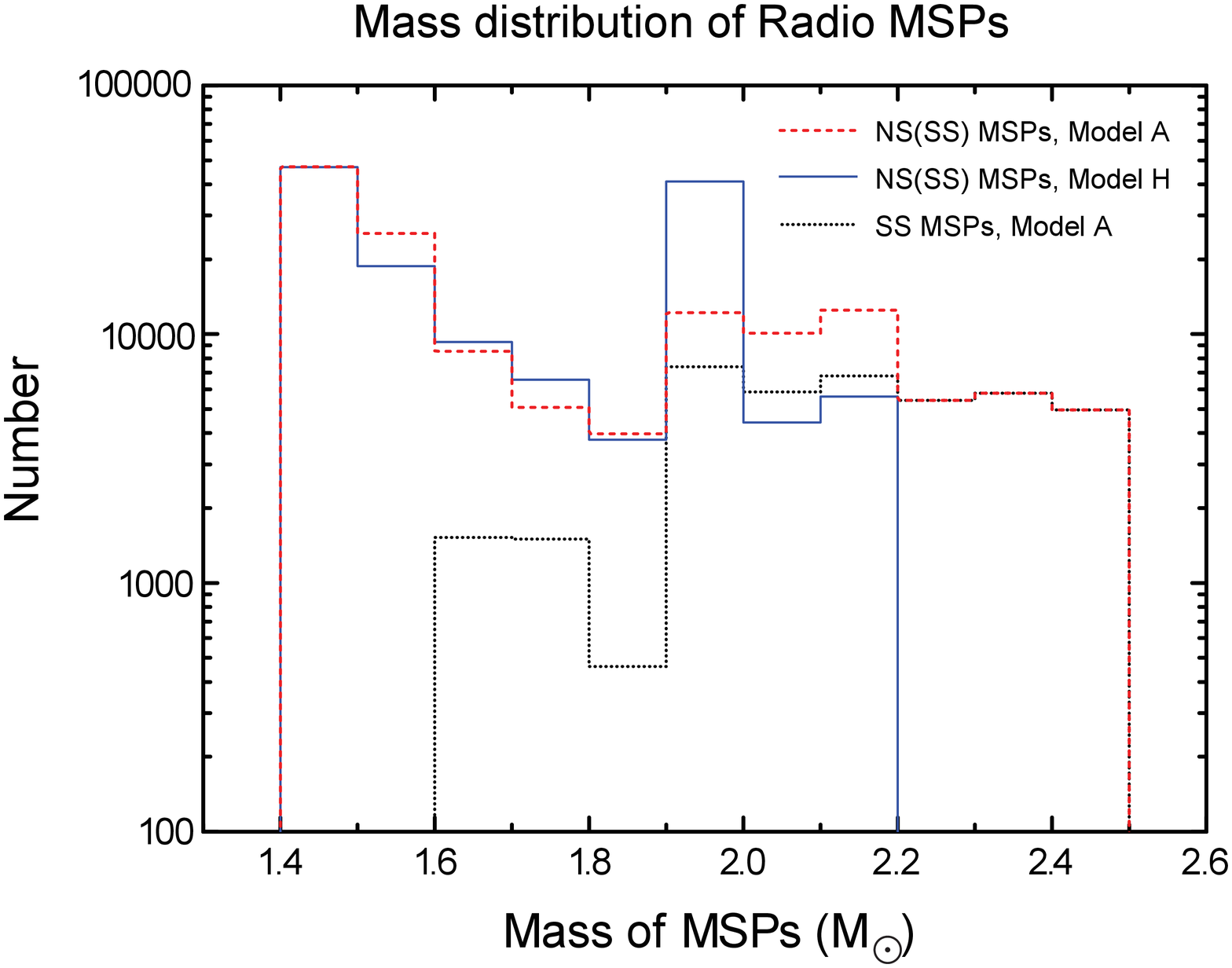}
\includegraphics[scale=0.34,trim={20 30 10 0},angle=0]{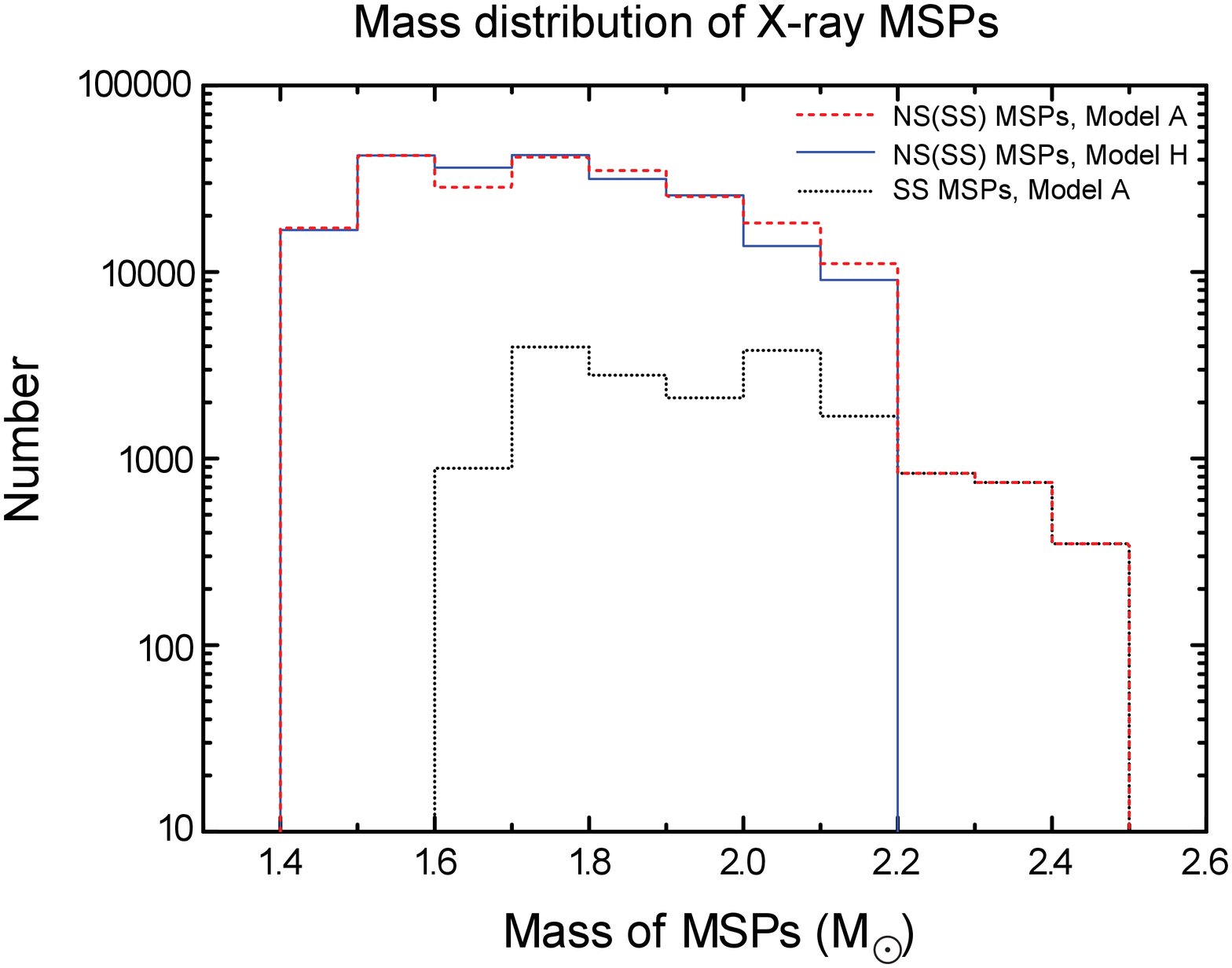}
\caption{Mass distribution of radio (top panel) and X-ray (bottom panel) MSPs in our simulated binaries in the Galaxy. The red dashed lines and black dashed lines represent the number of all binary MSPs (including NS and SS MSPs) predicted by Model A and H (twin of A, see the text 4.3, for detail), respectively.
The dotted lines correspond to SS MSPs predicted by Model A.}
\label{fig:distrAG}
\end{figure}

\begin{figure}
\centering
\includegraphics[scale=0.34,trim={20 30 0 0},angle=0]{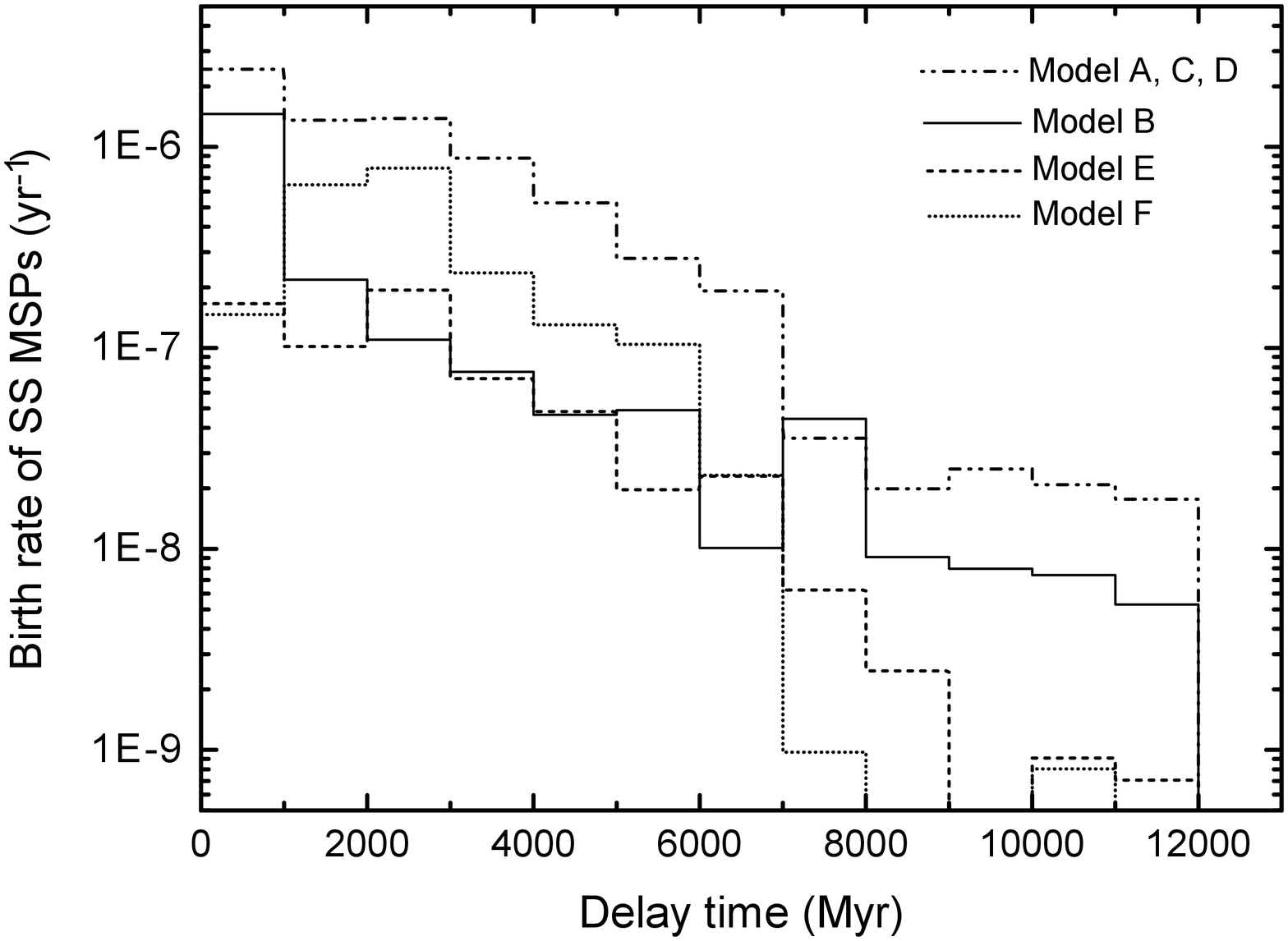}
\includegraphics[scale=0.34,trim={20 30 10 0},angle=0]{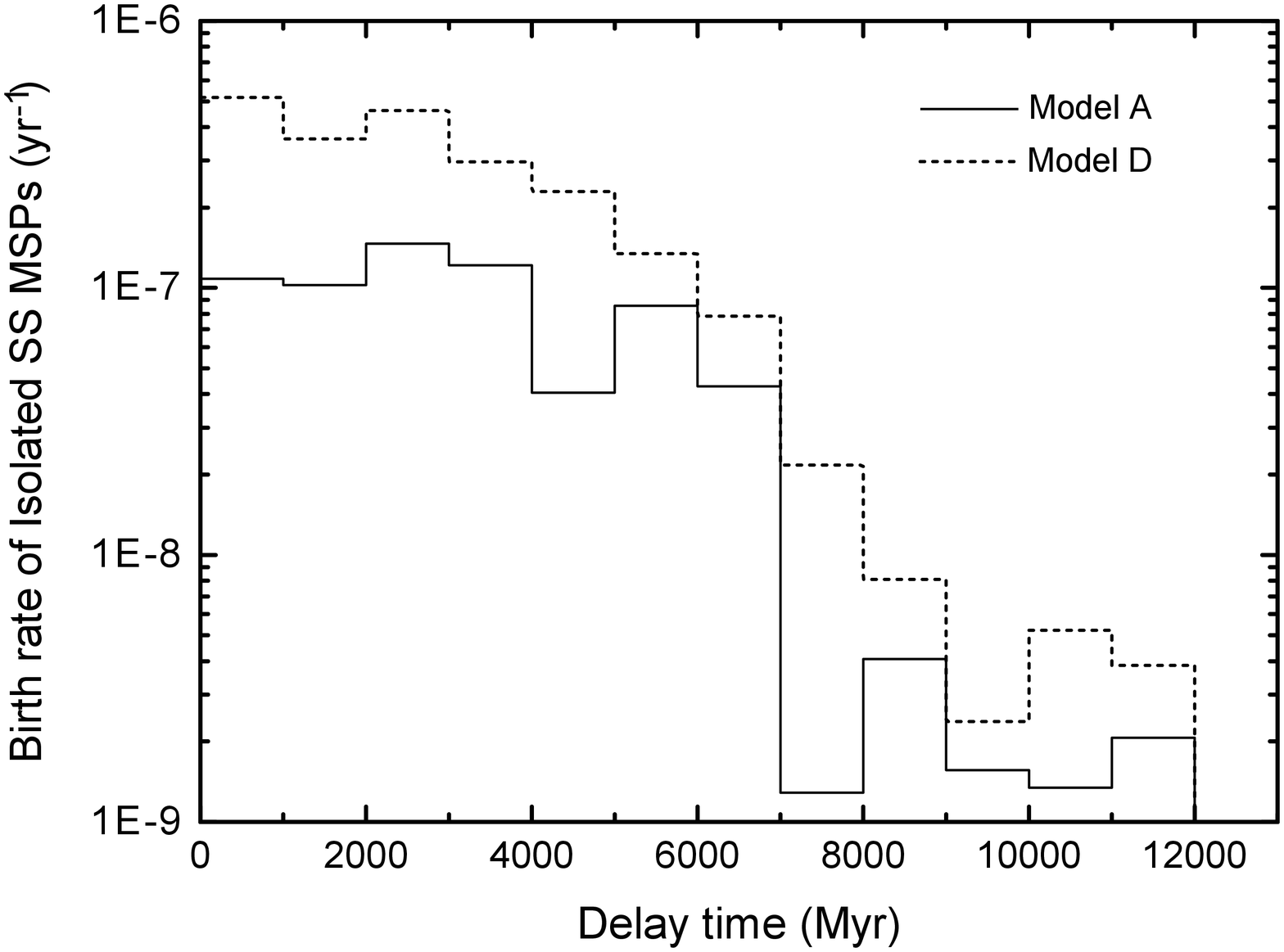}
\includegraphics[scale=0.34,trim={20 30 10 0},angle=0]{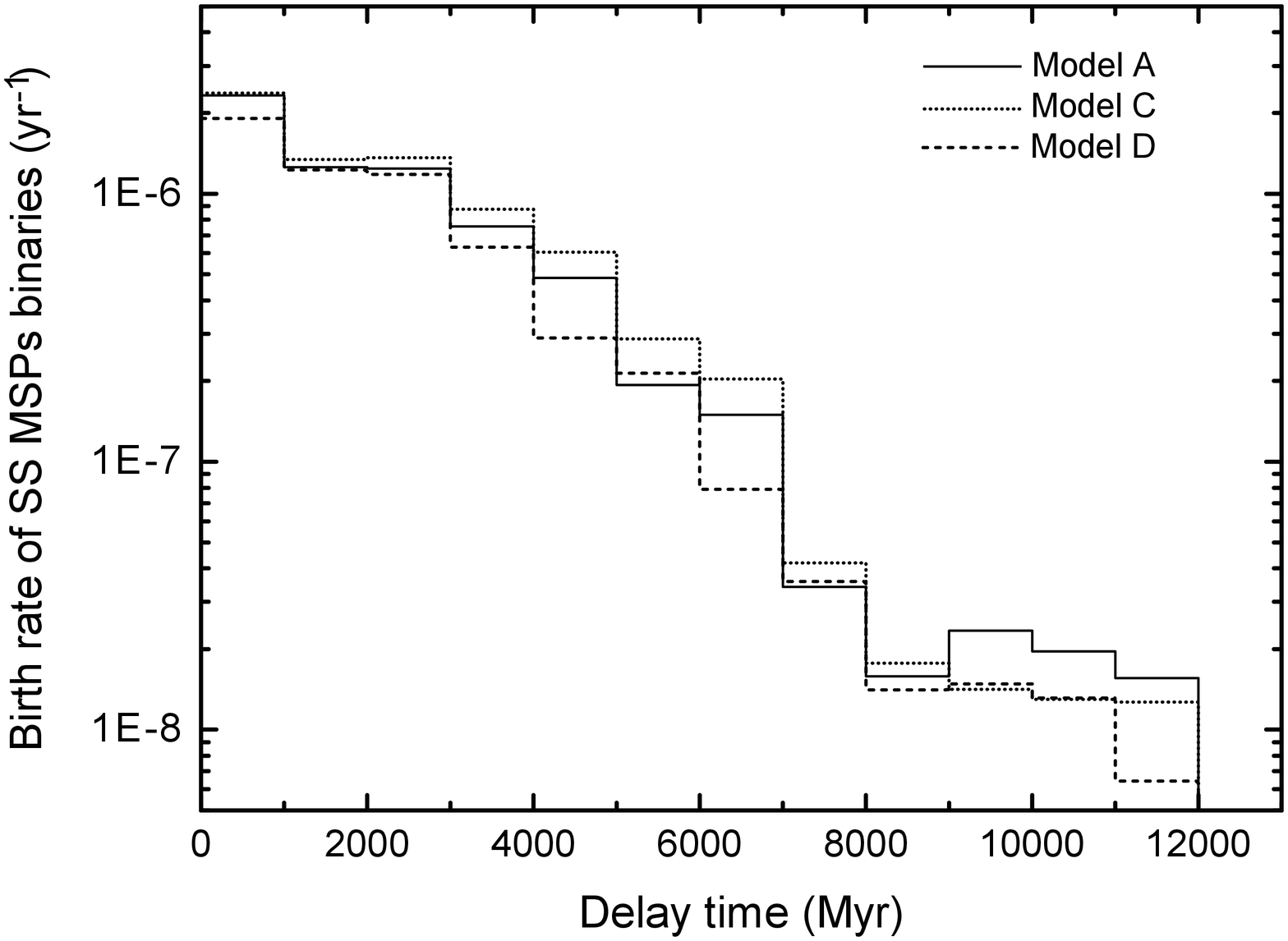}
\caption{Birth rate versus delay time. Top panel: influence of CE parameters on the birth rate of all SS MSPs;
middle panek: influence of PT kick velocity on the formation of isolated SS MSPs;
bottom panel: influence of PT kick velocity to the formation of SS binary MSPs. }
\label{fig:delay}
\end{figure}

\begin{figure}
\centering
\includegraphics[scale=0.34,trim={20 30 0 0},angle=0]{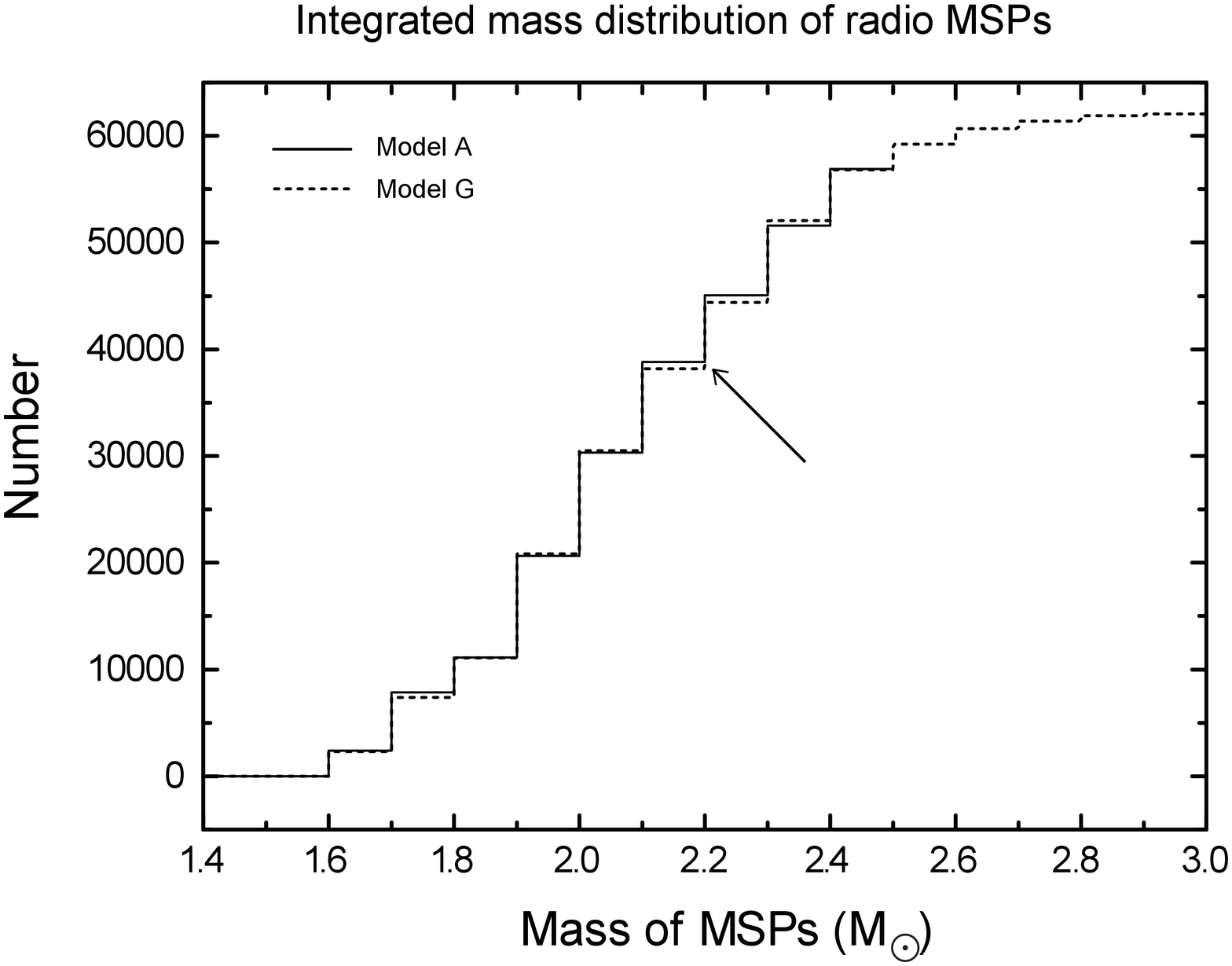}
\caption{Integrated number distribution of MSPs mass in radio binaries predicted by Model A (solid line) and G (dashed line), while the arrow indicates the cutoff with $M_{\rm SS, MAX}=2.2{\rm M}_{\odot}$ }
\label{fig:sumAH}
\end{figure}

\subsection{Predicted numbers and birth rates}
The predicted numbers and birth rates of different types of SS MSPs (radio, X-ray, with different companion types, and isolated MSPs)
in each model are summarized in Table 2. Some main results are summarized as follows:

(1) The birth rate and the total number of isolated SS MSPs in the Galaxy predicted by Model A
is $\sim6.6\times10^{-7}{\rm yr}^{-1}$ and $5878$, respectively, which is consistent with the lower limit predicted by \citet{lor95}, \citet{lyn98}, \citet{fer07} and \citet{sto07}.

(2) Models B, E and F yield a relatively low birth rate and total number, implying that binding energy parameter $\lambda$ and $\alpha_{\rm CE}$ play an important role in forming various SS MSPs. A higher $\alpha_{\rm CE}$ can prevent the binaries from coalescence during the CE phase, enhancing the birth rate of the post-CE binaries significantly \citep{liu06}.

(3) Comparison between the results of Models A, C, D indicates that a high kick velocity during PT can remarkably disrupt the binary systems, resulting in a high birth rate of isolated SS MSPs, while a low kick velocity with $\sigma_{\rm PT}=20{\rm km s}^{-1}$ can hardly disrupt the binary.

(4) Model G predicts similar results to Model A which indicates that, compared to other parameters, the influence of $M_{\rm SS,MAX}$ is minor.

(5) All models predict a considerable SS $+$ He WD binary MSPs,
while the predicted birth rates and numbers of other types of SS binary MSPs are much smaller than those of SS $+$ He WD binary MSPs.

\subsection{Evolutionary traces and initial parameters}
Fig. 1 shows the evolutionary traces of the mass of the NS and the donor star, orbital period, and eccentricity for three cases forming isolated MSP,
He WD$+$SS binary MSP, and CO WD$+$SS binary MSP in Model A.
It is easy to find the PT position because of the sudden mass decrease of the NS. In all three cases, the PT takes place during the RLOF stage.

In Fig. 2, the initial 3D parameter-space of $M_{\rm 1,i}$, $M_{\rm 2,i}$, and $P_{\rm orb,i}$ of the primordial binary systems that can evolve into isolated MSP,
 He WD $+$ MSP, and CO WD $+$ MSP in Model A are projected onto three planes.
The initial parameter distributions of the primordial binary systems forming isolated MSPs are very similar to those of He WD $+$ SS binary MSP
and their different evolutionary fates should originate from different kick velocities.
However, the primordial binary systems evolving into CO WD $+$ SS binary MSPs prefer to have a relatively heavy secondary star
which can be explained by the standard stellar evolutionary model.

\subsection{Mass distribution of Pulsars}

Considering its fast rotation, the mass of SS after PT may stop increasing as a result of the propeller effect.
The magnetosphere radius of the SS under the assumption of spherical accretion is
$r_{\rm m}=6.0\times10^6(B_9)^{4/7}|\dot{M}_{17}|^{-2/7}$~cm,
where $B_9$ is the surface magnetic field of SS in units of $10^9$~G
and $\dot{M}_{17}$ is the accretion rate of SS in units of $10^{17}{\rm~g~s}^{-1}$ \citep{GL79a, GL79b, liu11}.
The corotation radius of the SS can be estimated as:
\begin{equation}
r_{\rm co}=1.5\times10^6(\frac{M_{\rm SS}}{M_{\odot}})^{1/3}(\frac{P}{1{\rm ms}})^{2/3}{\rm ~cm}.
\end{equation}
The radius of the object after PT should decrease due to the mass loss.
According to the conservation of the magnetic flux, the surface magnetic field of the object would be enhanced,
while the spin period would decrease due to the conservation of the angular momentum.
As a result, the magnetosphere radius moves outside and the corotation radius moves inside.
Therefore, the propeller effect may possibly occur. Taking this possibility into
consideration, we also simulate Model H (the ¡¯twin¡¯ of Model
A, with all input parameters the same as for Model A) for which the mass
of SS will no longer increase after PT.

Fig. 3 summarizes the simulated mass distributions of various binary MSPs in the Galaxy.
The top and bottom panels show the mass distribution of radio (no mass accretion) and X-ray binary MSPs, respectively. The red dashed lines, and black dashed lines represent the number of all binary MSPs (including NS and SS MSPs) predicted by Model A and H, respectively. Since the further accretion after PT process ceases, the maximum mass of MSPs predicted by Model H is $2.2~\rm M_{\odot}$ (also see Equation 1).
The dotted lines correspond to SS MSPs predicted by Model A,
implying the contribution of core collapse NS obviously exceeds SS for MSPs with a mass of $1.6-2.2~\rm M_{\odot}$.
However, SS evolutionary channel provides the whole contribution for the radio MSPs with a mass exceeding $2.2~\rm M_{\odot}$. Recent observation revealed that there exists a heavy MSP with a mass of $2.1~\rm M_{\odot}$ \citep{yap19},
the NS-SS PT scenario may be responsible for the origin of such a heavy MSP.

\subsection{Influence of input parameters}

\subsubsection{CE parameters}
In Fig. 4, the birth rate of SS MSPs is shown as a function of the delay time between the formation of primordial binary systems and the PT. The influence of different CE parameters on the birth rate of all SS MSPs are shown in the top panel. It is clear that Models A, C and D with the same CE parameters lead to a large birth rate of SS MSPs.
\subsubsection{Kick velocity}
The middle and bottom panels indicate the influence of PT kick on the birth rate of isolates SS MSPs or SS binary MSPs, respectively. It is obvious that a higher kick velocity would easily lead to the disruption of binaries, yielding a relatively higher birth rate of isolated SS MSPs.
In addition, as shown in Table 2, when the kick velocity $\sigma_{\rm PT}=20\rm {~km~s}^{-1}$ (Model C), it is difficult to disrupt the binaries and produce isolated SS MSPs. As a result, there is no Model C in the middle panel of Fig. 4.
\subsubsection{Maximum mass of SS}
When the mass of SS in a binary reaches the maximum mass $M_{\rm SS,MAX}$ during the mass accretion, the SS binary MSP would evolve into a black hole binary and would not contribute to the birth rate of SS binary MSPs. Therefore, a lower maximum mass of SS should result in a lower birth rate of SS binary MSPs. For isolated SS MSPs with no mass transfer after PT, the birth rate is irrelevant to the maximum mass of SS. The minor difference of birth rates and number of isolated SS MSPs shown in Table 2 should arise from different random numbers. 

The integrated number distribution of binary radio MSPs (including NS and SS) predicted by Model A and G are shown in Fig. 5. Comparing the difference between models A and G in Table 2 and Fig. 5, the influence of the maximum mass on the birth rates and numbers of SS binaries with different companion types can be more or less neglected if we deduct the factor of the random numbers. The arrow in Fig. 5 indicates the cutoff point for the integrated mass distribution and total number of radio MSPs when $M_{\rm SS, MAX}=2.2{\rm M}_{\odot}$ (other input parameters are same to models A and G). 

\subsubsection{Accretion efficiency}
As shown in Table 2, the accretion efficiency $f_{\rm acc}$ can significantly influence the simulated results.
Generally, a higher accretion efficiency always results in higher birth rates and numbers of isolated SS MSPs. However, for binary SS MSPs with different companion types, the influence tendency is complicated. The accretion process would influence the orbital evolution of the binary system, resulting in different orbital periods and donor-star masses; hence, it would produce different birth rates and numbers of binary SS MSPs with different companion types.

\section{Discussion}

\subsection{Eccentric binary MSPs}
As mentioned above, in our simulation all PT processes take place during the RLOF
and the orbital circularization is very rapid due to the following mass exchange (as shown in the right middle and bottom panels of Fig. 1).
Therefore, it seems difficult to produce eccentric He WD $+$ MSP binaries, which are very rare in the Galaxy.
However, (1) the PT will occur if the mass of the NS is larger than $M_{\rm c}(\Omega)$ during the spin-down process \footnote{Since the magnetic fields of MSPs are very weak, we ignore the spin-down process that is due to the magnetic dipole radiation in the simulation.} and (2)
it is possible that the PT processes take place at the final stage of the mass transfer
(at the endpoint of RLOF or the time scale of the following mass transfer after PT is very short).
These terms indicate that the MSPs with He WD companions in eccentric orbit may originate from NS-SS PT.
\cite{jiang15} has already proposed a NS-SS PT model to account for the eccentric binary MSPs which arised from the sudden loss of the gravitational mass of the NS during the PT.

Apart from the NS-SS PT scenario, some other models can also account for the formation of eccentric binary MSPs.
\citet{ant14} suggests that the dynamical interaction between the binary and
a circumbinary disk (CB disk) could result in an eccentricity of $e\sim0.01-0.15$ for He WD binary MSPs with orbital periods between 15 and 50 days.

The scenario of delayed accretion-induced collapse (AIC) of accreting massive WD
first proposed by \citet{mic87} has been widely studied by different authors \citep{iva04, van97, xu09, hur10}.
\citet{fre14} suggest that the orbital eccentricity may be caused by a sudden mass loss during AIC.
Population-synthesis simulation given by \citet{che11} shows that the AIC scenario can also produce enough isolated MSPs in the Galaxy.
\citet{bar17} report the measurement of both the advance of periastron and the Shapiro delay for PSR J$1946+3417$
and obtain the mass of the pulsar, which is 1.828(22) M$_\odot$.
Since the Chandrasekhar mass of the WD is 1.4 M$_\odot$,
the mass of the MSP forming by the AIC channel should be $\sim 1.2~M_{\odot}$.
Even if we consider the collapse of a super-Chandrasekhar mass WD,
it is difficult to form such a heavy MSP \footnote{According to the simulation given by \citet{che09},
the massive WD exceeding $2.0~M_{\odot}$  cannot be produced by the mass accretion.}.

\subsection{MSPs with warm surfaces}
Some MSPs are reported to have relatively high surface temperatures, which are not consistent with their cooling evolution.
For example, the spectrum of PSR J0437-4715 and the WD companion can be fit
with surface temperatures of $\sim10^5 {\rm ~K}$ and $\sim4000 {\rm ~K}$, respectively \citep{dur12}.
The optical-far-UV spectrum of isolated MSP J2124-3358, observed by the {\textit Hubble Space Telescope,}
constrained its surface temperature at $\sim10^5 {\rm ~K}$ \citep{ran17}.
Considering the heating process during PT, the anomalous surface temperatures of PSR J0437-4715
and J2124-3358 can be interpreted by NS-SS PT scenario,
similarly to the double WD binary SDSS J125733.63+542850.5 studied by \citet{jiang18}.

\section{Summary}
In this work, we propose a NS-SS PT scenario to interpret the origin of isolated MSPs. Once the mass of the NS exceeds the maximum
mass due to the accretion, the NS-SS PT process occurs and a suitable kick would disrupt the binary, resulting in the birth of isolated MSPs. Employing the population-synthesis code, we simulate the evolution of $8\times10^6$ binary systems
for several models with different input parameters $\lambda, \alpha_{\rm CE}$, $\sigma_{\rm PT}$, $M_{\rm SS, MAS}$ and $f_{\rm acc}$.
The simulated results show that the NS-SS PT scenario with a kick velocity of $\sigma_{\rm PT}=60 {\rm ~km~s}^{-1}$
can produce a considerable isolated MSPs, which is approximately in agreement with the predictions given by \citet{lor95,lyn98,fer07,sto07}. Meanwhile, disrupted binary MSPs with He WDs should be responsible for the origin of isolated low mass He WDs, which cannot evolve from the normal single star evolutionary channel \citep{wan09, zor17}.  In the present scenario, most donor stars, companions of the progenitors of isolated SS MSPs, would
evolve into He WDs in the Hubble time, hence, the upper limit of birth rate of isolated low-mass He WDs via NS-SS PT should be similar to that of isolated MSPs.
In addition, the scenario also predicts considerable He WD $+$ SS binary MSPs
and the mass distribution of MSPs can be used to check the current scenario.

At present, the predicted NS-SS PT event has never been confirmed by
observation. In principle, current transient surveys may hint at some SS candidates. An NS-SS PT event could be observed as short $\gamma$-ray burst \citep{che96} or quark-nova \citep{ouy02}, releasing $10^{53}{\rm ~ergs}$ in a short timescale (less than one second). Recently, \cite{yu19} predicted that the accretion-induced collapse (AIC) of white dwarfs should be associated with recently discovered fast-evolving luminous transients \citep{yu15}, which has been observed in all of
the optical, soft, and hard X-ray bands. If the NS-SS PT process produces a highly magnetized MSPs, the observed phenomenon should be similar to the AIC. Certainly, the mass of SS MSPs should be greater than that of MSPs forming by AIC.

\begin{acknowledgements}
We thank the referee for a very careful reading and
comments that have led to the improvement of the manuscript. This work was supported by the CAS 'Light of West China' Program (Grants No. 2018-XBQNXZ-B-022) and the National Natural Science Foundation of China (Grant Nos. 11573016, 11733009, 11773015, 11333004, U1731103, 11463004 and 11605110), the National Key Research and Development Program of China (Grant No. 2016YFA0400803), the Scientific Research Fund of Hunan Provincial Education Department (Grant Nos. 16B250 and 16C1531), and the Program for Innovative Research Team (in Science and Technology) at the University of Henan Province.
\end{acknowledgements}

\end{document}